\documentstyle[12pt,psfig]{article}
\catcode`\@=11
\def\marginnote#1{}
\newcount\hour
\newcount\minute
\newtoks\amorpm
\hour=\time\divide\hour by60
\minute=\time{\multiply\hour by60 \global\advance\minute by-
\hour}
\edef\standardtime{{\ifnum\hour<12 \global\amorpm={am}%
    \else\global\amorpm={pm}\advance\hour by-12 \fi
    \ifnum\hour=0 \hour=12 \fi
    \number\hour:\ifnum\minute<100\fi\number\minute\the\amorpm}}
\edef\militarytime{\number\hour:\ifnum\minute<100\fi\number\minute}
\def\draftlabel#1{{\@bsphack\if@filesw {\let\thepage\relax
  \xdef\@gtempa{\write\@auxout{\string
    \newlabel{#1}{{\@currentlabel}{\thepage}}}}}\@gtempa
    \if@nobreak \ifvmode\nobreak\fi\fi\fi\@esphack}
     \gdef\@eqnlabel{#1}}
\def\@eqnlabel{}
\def\@vacuum{}
\def\draftmarginnote#1{\marginpar{\raggedright\scriptsize\tt#1}}
\def\draft{\oddsidemargin -.5truein
        \def\@oddfoot{\sl preliminary draft \hfil
        \rm\thepage\hfil\sl\today\quad\militarytime}
        \let\@evenfoot\@oddfoot \overfullrule 3pt
        \let\label=\draftlabel
        \let\marginnote=\draftmarginnote

\def\@eqnnum{(\theequation)\rlap{\kern\marginparsep\tt\@eqnlabel}%
\global\let\@eqnlabel\@vacuum}  }
\def\preprint{\twocolumn\sloppy\flushbottom\parindent 1em
        \leftmargini 2em\leftmarginv .5em\leftmarginvi .5em
        \oddsidemargin -.5in    \evensidemargin -.5in
        \columnsep 15mm \footheight 0pt
        \textwidth 250mmin      \topmargin  -.4in
        \headheight 12pt \topskip .4in
        \textheight 175mm
        \footskip 0pt

\def\@oddhead{\thepage\hfil\addtocounter{page}{1}\thepage}
        \let\@evenhead\@oddhead \def\@oddfoot{} \def\@evenfoot{}
}
\def\titlepage{\@restonecolfalse\if@twocolumn\@restonecoltrue\onecolumn
     \else \newpage \fi \thispagestyle{empty}\c@page\z@
        \def\thefootnote{\fnsymbol{footnote}} }
\def\endtitlepage{\if@restonecol\twocolumn \else  \fi
        \def\thefootnote{\arabic{footnote}}
        \setcounter{footnote}{0}}  %\c@footnote\z@ }
\catcode`@=12
\relax
\def\be{\begin{equation}}
\def\ee{\end{equation}}
\def\bea{\begin{eqnarray}}
\def\eea{\end{eqnarray}}
\def\simlt{\stackrel{<}{{}_\sim}}

\def\NPB#1#2#3{{\it Nucl.~Phys.} {\bf{B#1}} (19#2) #3}
\def\PLB#1#2#3{{\it Phys.~Lett.} {\bf{B#1}} (19#2) #3}
\def\PRD#1#2#3{{\it Phys.~Rev.} {\bf{D#1}} (19#2) #3}
\def\PRL#1#2#3{{\it Phys.~Rev.~Lett.} {\bf{#1}} (19#2) #3}

\def\PTP#1#2#3{{\it Prog.~Theor.~Phys.} {\bf#1}  (19#2) #3}

\def\mst1{m_{\widetilde{t}_1}}
\def\mst2{m_{\widetilde{t}_2}}
\def\mst12{m_{\widetilde{t}_{1,2}}}

\def\msb1{m_{\widetilde{b}_1}}
\def\msb2{m_{\widetilde{b}_2}}
\def\msb12{m_{\widetilde{b}_{1,2}}}

\def\mtilde2{\widetilde{m}^{2}}

\relax
\begin{document}
\setlength{\baselineskip}{3.0ex}
\begin{titlepage}
%\phantom{bla}
\begin{flushright}
IEM-FT-114/95 \\
hep--ph/9509385 \\
\end{flushright}
\vspace{3.5cm}
%\vskip 0.3in
\begin{center}{\large\bf
STATUS OF EFFECTIVE POTENTIAL CALCULATIONS
\footnote{Based on talk given at the {\it SUSY-95 International
Workshop on Supersymmetry and Unification of Fundamental
Interactions}, Palaiseau, 15-19 May 1995. }  } \\
\vspace*{6.0ex}
{\large M. Quir\'os}
\footnote{On leave of absence from Instituto
de Estructura de la Materia, CSIC, Serrano 123, 28006-Madrid,
Spain. Work supported in part by
the European Union (contract CHRX-CT92-0004) and
CICYT of Spain
(contract AEN94-0928).}\\
\vspace*{1.5ex}
{\large\it CERN, TH Division, CH--1211 Geneva 23, Switzerland}\\
\end{center}
\vspace{1cm}

\vskip2.5cm
\begin{center}
{\bf Abstract}
\end{center}
%\begin{quote}
%\vbox{ \baselineskip 14pt
We review various effective potential methods which have been useful
to compute the Higgs mass spectrum and couplings of the minimal
supersymmetric standard model. We compare results where all-loop
next-to-leading-log corrections are resummed  by the
renormalization group, with those where just the leading-log
corrections are kept. Pole masses are obtained from
running masses by addition of convenient self-energy diagrams.
Approximate analytical expressions are worked out, providing an
excellent approximation to the numerical results which include all
next-to-leading-log terms. An appropriate
treatment of squark decoupling allows to consider large values of
the stop and/or sbottom mixing parameters and thus fix a reliable upper
bound on the mass of the lightest CP-even Higgs boson mass.
%\end{quote}
%}
%\vskip1.cm

\vspace{3cm}
\begin{flushleft}
IEM-FT-114/95\\
September 1995 \\
\end{flushleft}

\end{titlepage}
\newpage
\section{Introduction}

The {\bf effective potential} methods to compute the (radiatively
corrected) Higgs mass spectrum in the Minimal Supersymmetric
Standard Model (MSSM) are useful since they allow to {\bf resum}
(using Renormalization Group (RG) techniques) leading-log (LL),
next-to-leading-log (NTLL),..., corrections to {\bf all orders}
in perturbation theory. These methods~\cite{Effpot,EQ}, as well as the
diagrammatic methods~\cite{Diagram} to compute the Higgs mass spectrum
in the MSSM, were first developed in the early nineties.

Effective potential methods are based on the {\bf run-and-match}
procedure by which all dimensionful and dimensionless couplings
are running with the RG scale, for scales greater than the
masses involved in the theory. When the RG scale
equals a particular mass threshold, heavy fields decouple,
eventually leaving threshold effects in order to match the
effective theory below and above the mass threshold. For
instance, assuming a common soft supersymmetry breaking mass
for left-handed and right-handed stops and sbottoms,
$M_S\sim m_Q\sim m_U\sim m_D$, and assuming for the top-quark mass,
$m_t$, and for the CP-odd Higgs mass, $m_A$, the range
$m_t\leq m_A\leq M_S$, we have: for scales $Q\geq M_S$, the MSSM, for
$m_A\leq Q\leq M_S$ the two-Higgs doublet model (2HDM), and for
$m_t\leq Q\leq m_A$ the Standard Model (SM). Of course there are
thresholds effects at $Q=M_S$ to match the MSSM with the 2HDM, and
at $Q=m_A$ to match the 2HDM with the SM.

The neutral Higgs sector of the MSSM contains,
on top of the CP-odd Higgs $A$, two CP-even Higgs mass
eigenstates, $H_h$ (the heaviest one) and $H$ (the lightest one).
It turns out that the larger
$m_A$ the heavier the lightest Higgs $H$. Therefore the case
$m_A\sim M_S$ is, not only a great simplification since the effective
theory below $M_S$ is the SM, but also of great interest, since it
provides the upper bound on the mass of the lightest Higgs
(which is interesting for phenomenological purposes, e.g. at
LEP~2). In this case the threshold correction at $M_S$ for the SM
quartic coupling $\lambda$ is:
\be
\label{threshold}
\Delta_{\rm th}\lambda=\frac{3}{16\pi^2}h_t^4
\frac{X_t^2}{M_S^2}\left(2-\frac{1}{6}\frac{X_t^2}{M_S^2}\right)
\ee
where $h_t$ is the SM top Yukawa coupling and
$X_t=(A_t-\mu/\tan\beta)$ is the mixing in the stop mass
matrix, the parameters $A_t$ and $\mu$ being the trilinear
soft-breaking coupling in the stop sector and the supersymmetric
Higgs mixing mass, respectively. The maximum of
(\ref{threshold}) corresponds to $X_t^2=6 M_S^2$ which provides
the maximum value of the lightest Higgs mass: this case will be
referred to as the case of maximal mixing.

\section{Leading-log vs Next-to-leading-log results}

Recent effective potential studies~\cite{NTLL} have proved that
the L-loop improved effective potential, with (L+1)-loop RG
equations is exact to Lth-to-leading-log order. In particular
the case L=0 (tree level potential improved with one-loop RG)
describes the LL approximation, and the case L=1 (one-loop
effective potential improved with two-loop RG) describes the
NTLL approximation. In particular, the effective potential in
the NTLL approximation is expected to be highly scale
independent. But, can we quantify the scale independence of it?

We have minimized the effective potential at the scale $t^*$
such that~\cite{CEQR}
\be
\label{tstar}
\frac{d}{dt}\left.\frac{\Phi_{\rm
min}(t)}{\xi(t)}\right|_{t=t^*}=0
\ee
%
%%%%%%%%%%%%%%%%%%%%%%%%%%%%%%%%%%%%%%%%%%%%%
\begin{figure}[htb]
%\psdraft
\centerline{
\psfig{figure=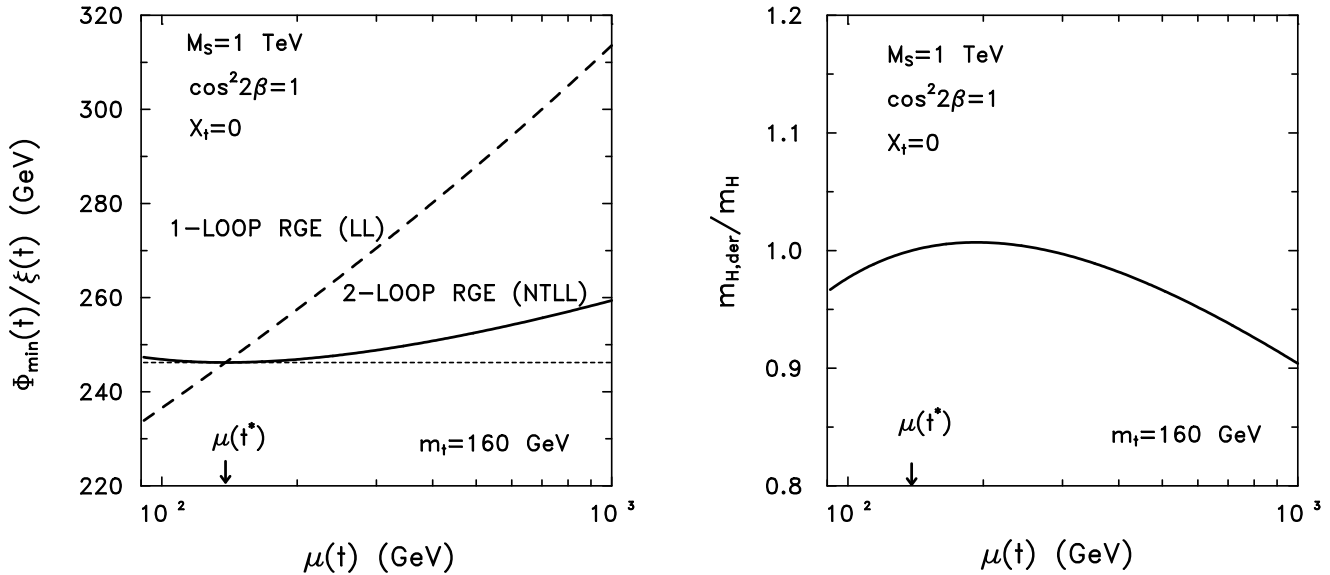,height=9.5cm,bbllx=9.5cm,bblly=1.cm,bburx=19.cm,bbury=14cm}}
\caption{{\bf Left panel:} Plot of $\phi_{\rm min}(t)/\xi(t)$
as a function of $\mu(t)$ in the
LL (dashed line) and NTLL (solid line) approximation. The dotted line shows the
would-be scale independent result.
{\bf Right panel:} Plot of $m_{H,{\rm der}}(t)/m_H(t)$
as a function of $\mu(t)$. In both cases
$m_t=160$ GeV and the supersymmetric parameters
are $M_S=1$ TeV, $X_t=0$ and $\tan\beta\gg 1$.}
\end{figure}
%%%%%%%%%%%%%%%%%%%%%%%%%%%%%%%%%%%%%%%%%%%%%%%
%%%%%%%%%%%%%%%%%%%%%%%%%%%%%%%%%%%%%%%%%%%%%
\begin{figure}[htb]
%\psdraft
\centerline{
\psfig{figure=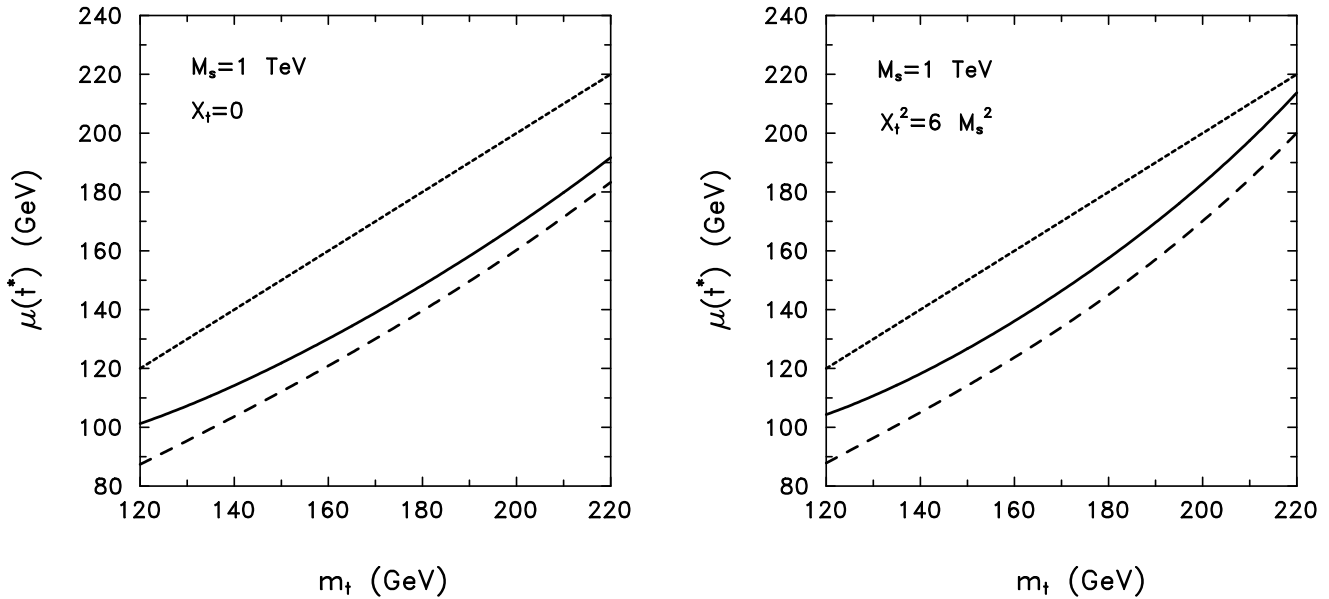,height=9.5cm,bbllx=9.5cm,bblly=1.cm,bburx=19.cm,bbury=14cm}}
\caption{Plot of $\mu(t)^*$ as a function of $m_t$ for
$\tan\beta\gg 1$ (solid lines),
$\tan\beta=1$ (dashed lines), and supersymmetric parameters $M_S=1$ TeV,
$X_t=0$ (left panel) and  $X_t^2=6 M_S^2$ (right panel).
The dotted lines correspond
to $\mu(t^*)=m_t$.}
\end{figure}
%%%%%%%%%%%%%%%%%%%%%%%%%%%%%%%%%%%%%%%%%%%%%%%
%%%%%%%%%%%%%%%%%%%%%%%%%%%%%%%%%%%%%%%%%%%%%
\begin{figure}[htb]
%\psdraft
\centerline{
\psfig{figure=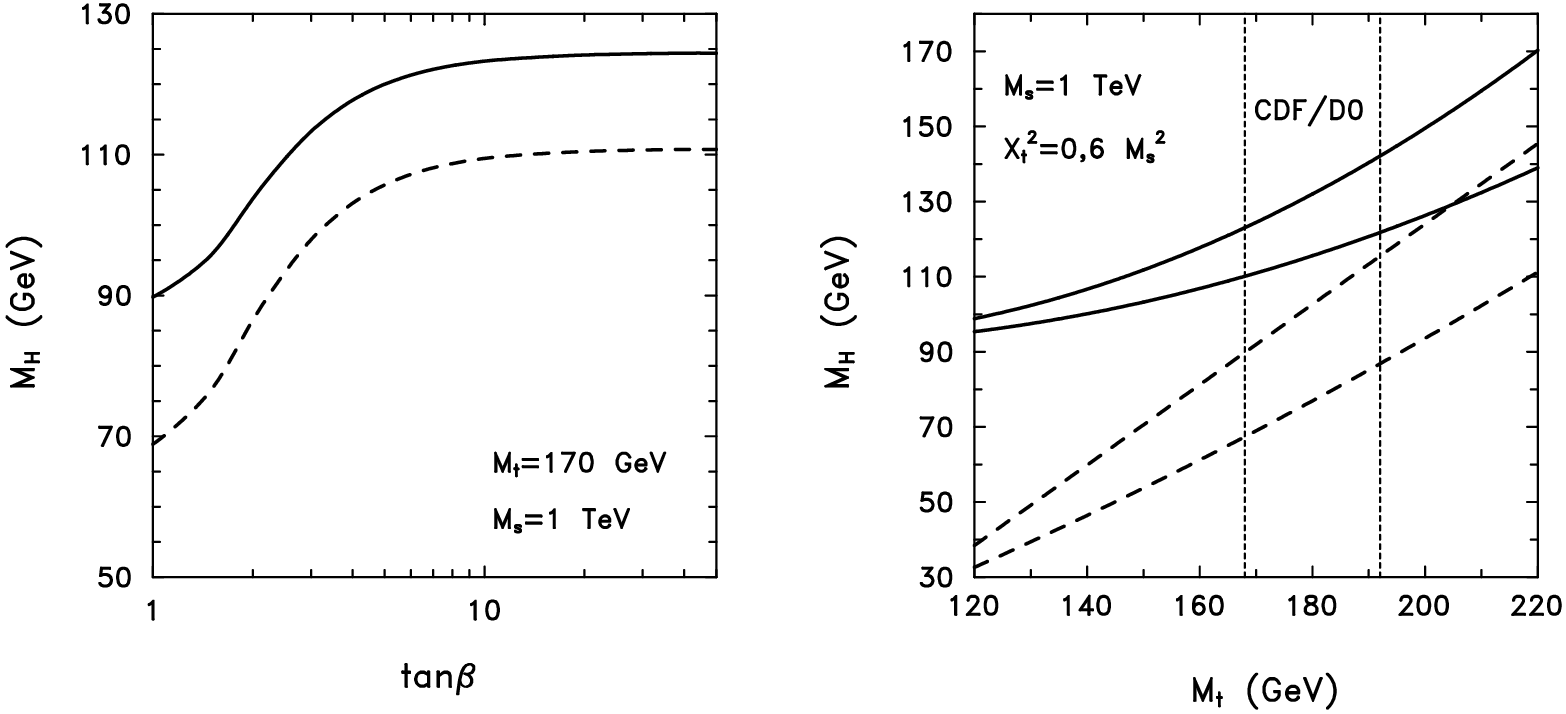,height=9.5cm,bbllx=9.5cm,bblly=1.cm,bburx=19.cm,bbury=14cm}}
\caption{{\bf Left panel:} Plot of $M_H$ as a function of
$\tan\beta$ for $M_t=170$ GeV
and $M_S=1$ TeV. The solid (dashed) curve corresponds
to the  $X_t^2=6M_S^2$ ($X_t=0$) case.
{\bf Right panel:} Plot of $M_H$ as a function of $M_t$ for $\tan\beta\gg 1$
(solid lines), $\tan\beta=1$ (dashed lines), and $X_t^2=6 M_S^2$ (upper set),
$X_t=0$ (lower set). The experimental band from the CDF/D0 detection is also
indicated.}
\end{figure}
%%%%%%%%%%%%%%%%%%%%%%%%%%%%%%%%%%%%%%%%%%%%%%%
where $\xi(t)$ is the anomalous dimension of the Higgs field.
In fact the scale dependence is measured by the quantity
$\Phi_{\rm min}(t)/\xi(t)$, which is constant for a scale
invariant theory. We have plotted it versus the RG scale $\mu(t)$
in Fig.~1 (left panel) where we can see that the best
minimization scale $\mu(t^*)$ is around the value of the running
top-quark mass and that minimizing the effective potential at
the high scale $M_S$ produces a departure of the scale
invariance of $\Phi_{\rm min}(t)/\xi(t)$ of order 10\%. This measure of
scale invariance is associated with the minimum of the
potential, i.e. with the first derivative. An independent
measure, associated with the second derivative of the effective
potential, is given by the ratio $m_{H,der}/m_H$, where
$m_{H,der}$ is the second derivative of the potential at the
scale $\mu(t)$ and $m_H$ the second derivative at the
minimization scale $\mu(t^*)$, run to the scale $\mu(t)$ with
the anomalous dimension of the Higgs field.
The corresponding
plot is shown in Fig.~1 (right panel) where we can see a similar
tendency in the sense that the ratio $(m_{H,der}-m_H)/m_H$ is ${\cal
O}(10)\%$ for large values of the scale $\mu(t)$.
This feature of $\mu(t^*)$ being close to the top-quark running
mass on shell (i.e. $m_t(Q=m_t)=m_t$) remains for all values of
the top-quark mass and supersymmetric parameters, as in Fig.~2.

We also have defined physical (pole) masses for the top quark
and the Higgs boson. For the top-quark we have included
the usual one-loop
QCD corrections, and for the Higgs boson one-loop electroweak
corrections as
\be
\label{higgspole}
M_H^2=m_H^2(t)+{\rm Re}\left[\Pi_{HH}(M_H^2)-\Pi_{HH}(0)\right]
\ee
where $m_H(t)$ is the running Higgs mass and $\Pi_{HH}$ the
Higgs self-energies which can be found in Ref.~\cite{CEQR}. We
have plotted in Fig.~3 $M_H$ as a function of $\tan\beta$ (left
panel) and as a function of $M_t$ (right panel). From Fig.~3 we
can see that the present experimental band from CDF/D0 for the
top-quark mass requires $M_H\simlt 140$ GeV, while if we fix
$M_t=170$ GeV, then the upper bound $M_H\simlt 125$ GeV
follows. It goes without saying
that these figures are extremely relevant for MSSM Higgs searches
at LEP~2.

\section{Comparison with other approaches}

It is worth at this point to compare our NTLL calculation with
other similar calculations aiming to evaluate two-loop
corrections to the lightest Higgs mass in the MSSM. For this
purpose we have plotted in Fig.~4 $M_H$ as a function of the
minimization scale $\mu(t^*)$ for different values of
supersymmetric parameters, and for the LL (dashed lines) and
NTLL (solid lines) approximations. We can see that our
definition of pole mass (\ref{higgspole}) provides a highly
scale independent Higgs mass in the NTLL approximation.
Moreover, as was the case in Fig.~1 (left panel), the LL
approximation exhibits a strong dependence with the minimization
scale.

%%%%%%%%%%%%%%%%%%%%%%%%%%%%%%%%%%%%%%%%%%%%%
\begin{figure}[htb]
%\psdraft
\centerline{
\psfig{figure=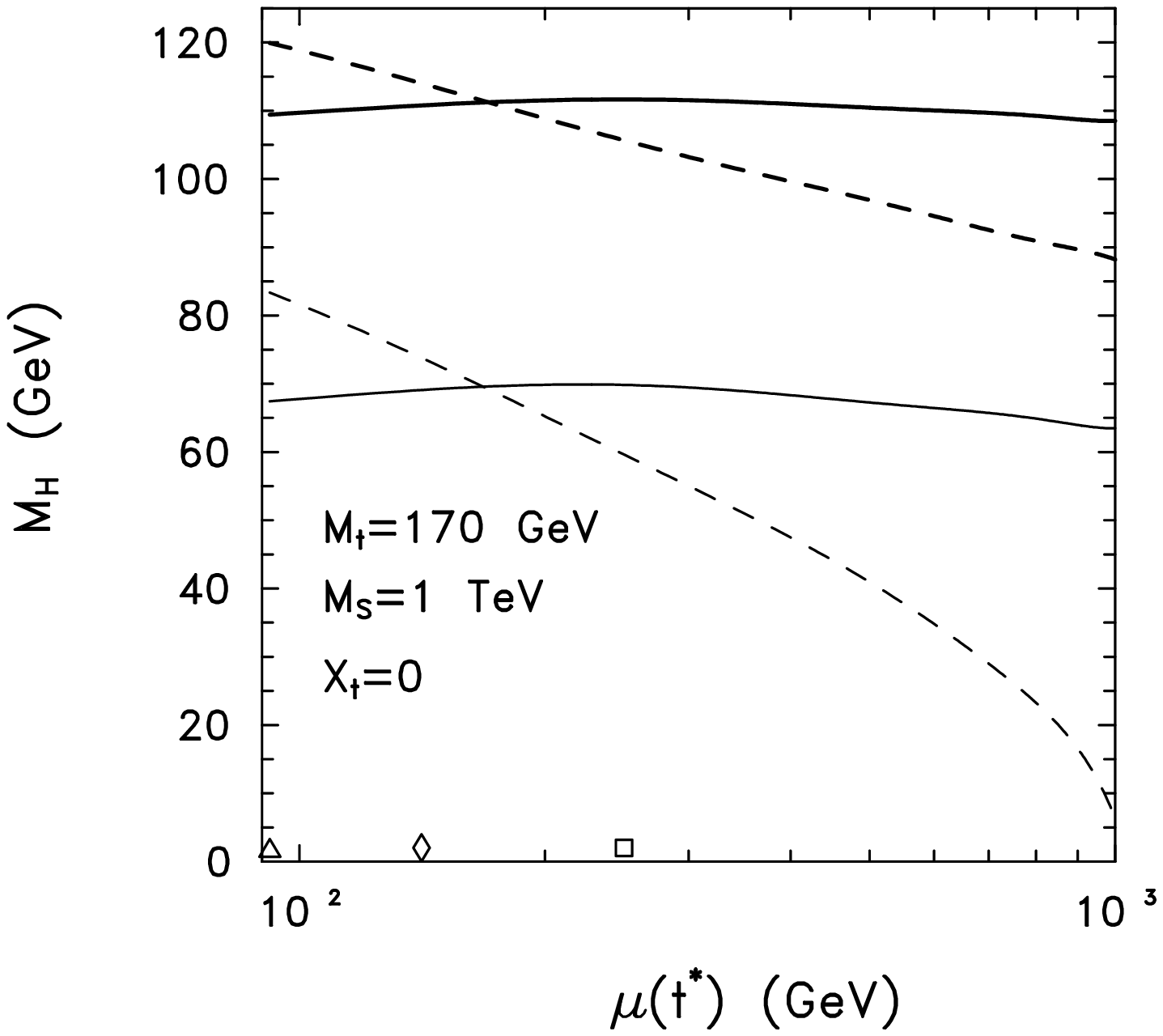,height=9.cm,bbllx=11.cm,bblly=3.cm,bburx=20.cm,bbury=16cm}}
\caption{Plot of $M_H$ as a function of $\mu(t^*)$
for $M_t=170$ GeV, $M_S=1$ TeV and $X_t=0$.
The solid (dashed) curves correspond to the NTLL (LL) approximation,
and the upper (lower) set to $\tan\beta\gg 1$ ($\tan\beta=1$).}
\end{figure}
%%%%%%%%%%%%%%%%%%%%%%%%%%%%%%%%%%%%%%%%%%%%%%%

$\Box$  Kodaira, Yasui and Sasaki~\cite{KYS} have used an
effective potential method neglecting gauge couplings in the
one-loop correction to the effective potential ($g=g'=0$) and
neglecting the contribution of the stop mixing to the threshold
corrections ($X_t=0$). Furthermore they have not introduced any wave
function renormalization for the Higgs
and thus have worked with running Higgs
masses. Finally they have chosen as minimization scale
$\mu(t^*)=v=246.22$ GeV and claimed that the difference between
the NTLL and the LL approximations is positive and sizeable,
unlike a previous result in Ref.~\cite{EQ} where it was claimed
it to be negative and small. A quick glance at Fig.~4 allows to
understand these seemingly contradictory results. We have
plotted with a $\Box$ the minimization scale chosen in
Ref.~\cite{KYS} and with a $\Diamond$ our choice. Had we chosen
as minimization scale the $\Box$ we had found (even if our
analysis includes gauge coupling effects and considers pole
masses) that (NTLL--LL) is positive and sizeable. However our
choice of the minimization scale at the $\Diamond$ position
suggests (NTLL--LL) negative and tiny. Since we made our choice
of the minimization scale such that the effective potential is
as scale independent as possible we could say that the latter
statement is the correct one. However, since we have defined pole
(scale independent) masses for the Higgs boson (unlike in
Ref.~\cite{KYS}) in the NTLL approximation, and the LL
approximation is strongly scale dependent, the quantity
(NTLL--LL) is scale dependent and then meaningless to qualify the
goodness of the approximation.

$\triangle$ Langacker and Polonsky~\cite{LP} work in the LL
approximation and fix as minimization scale $\mu(t^*)=M_Z$. They
find larger values than those in Fig.~3. We have plotted in
Fig.~4 their minimization scale with a $\triangle$. We can
easily see that for that scale the two-loop corrections that
they disregard are large and negative. Had they evaluated them
they should have found similar results to ours.

$\Diamond$ Finally, Hempfling and Hoang~\cite{HH} use
diagrammatic and effective potential approaches to evaluate the
lightest Higgs mass at two-loop order. They use various
approximations, as e.g. $g=g'=0$ in the effective potential and
only deal with the case of zero stop mixing, $X_t=0$. Their
results contain LL and NTLL corrections to one- and two-loop
orders, while our LL and NTLL corrections are resummed to
all-loop by the RG. However, in the case of no mixing, where we
can compare both approaches, our results agree within $\sim$2-3
GeV, which we consider as fully satisfactory, given all their
simplifications and the fact that this corresponds to the
uncertainty of our own calculation, due to the tiny scale
dependence of the effective potential and pole masses, and other
small effects as, e.g., the possible presence of light charginos and
neutralinos.

\section{An analytical approximation}

We have seen from Figs.~1 and 2, and previous considerations that,
since radiative corrections are minimized for scales $Q\sim m_t$,
when the LL RG improved Higgs mass expressions~\cite{LLRG} are
evaluated at the top-quark mass scale, they reproduce the NTLL value
with a high level of accuracy, for any value of $\tan\beta$ and the
stop mixing parameters~\cite{CEQW}
\be
\label{relmasas}
m_{H,LL}(Q^2\sim m_t^2)\sim m_{H,NTLL}.
\ee
Based on the above observation, we can work out a very accurate
analytical approximation to $m_{H,NTLL}$ by just keeping two-loop
LL corrections at $Q^2=m_t^2$, i.e. corrections of order $t^2$, where
$t=\log(M_S^2/m_t^2)$.

Again the case $m_A\sim M_S$ is the simplest, and very illustrative,
one. We have found~\cite{CEQW,HHH} that, in the absence of mixing
(the case $X_t=0$) two-loop corrections resum in the one-loop
result shifting the energy scale from $M_S$ (the tree-level scale)
to $\sqrt{M_S\; m_t}$. More explicitly,
\be
\label{resum}
m_H^2=M_Z^2 \cos^2 2\beta\left(1-\frac{3}{8\pi^2}h_t^2\; t\right)
+\frac{3}{2\pi^2 v^2}m_t^4(\sqrt{M_S m_t}) t
\ee
where $v=246.22$ GeV.

%%%%%%%%%%%%%%%%%%%%%%%%%%%%%%%%%%%%%%%%%%%%%%%
\begin{figure}[htb]
%\psdraft
\centerline{
\psfig{figure=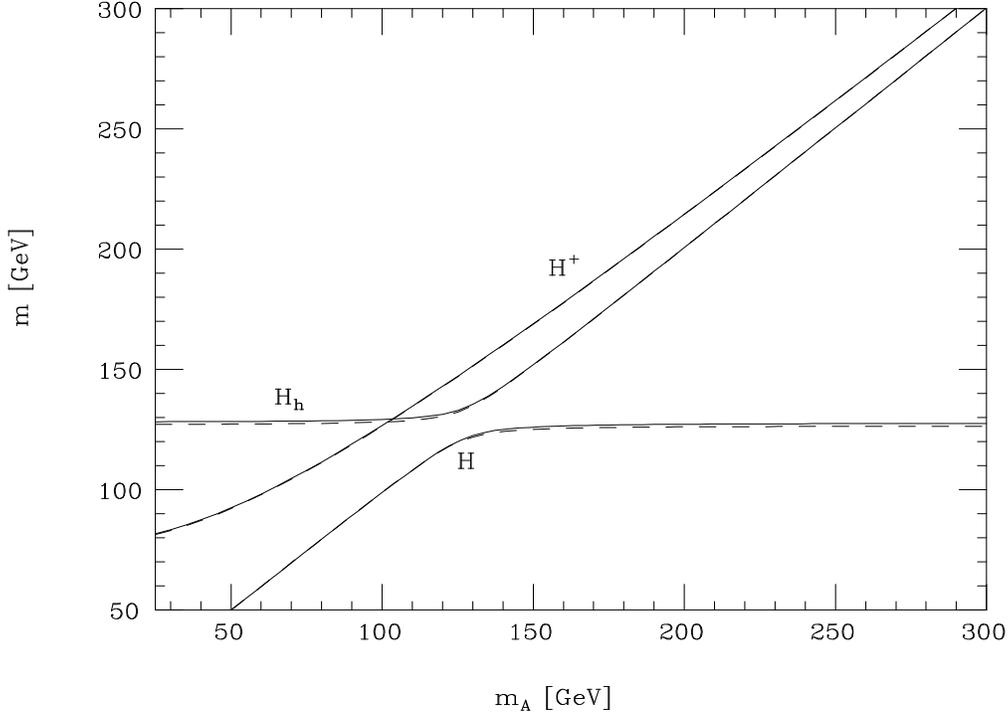,width=17cm,height=12cm,angle=90} }
\caption[0]
{The neutral ($H_h,H$) and charged ($H^+$) Higgs mass spectrum
as a function of the CP-odd Higgs mass $m_A$ for
a physical top-quark mass $M_t =$ 175 GeV and $M_S$ = 1 TeV, as
obtained from the one-loop improved RG evolution
(solid lines) and the analytical formulae (dashed lines).
All sets of curves correspond to
$\tan \beta=$ 15 and large squark mixing, $X_t^2 = 6 M_S^2$
($\mu=0$).}
\end{figure}
%%%%%%%%%%%%%%%%%%%%%%%%%%%%%%%%%%%%%%%%%%%%%%%%%%%%

In the presence of mixing ($X_t\neq 0$), the run-and-match procedure
yields an extra piece in the SM effective potential
$\Delta V_{\rm th}[\phi(M_S)]$ whose second derivative gives an
extra contribution to the Higgs mass, as
\be
\label{Deltathm}
\Delta_{\rm th}m_H^2=\frac{\partial^2}{\partial\phi^2(t)}
\Delta V_{\rm th}[\phi(M_S)]=
\frac{1}{\xi^2(t)}
\frac{\partial^2}{\partial\phi^2(t)}
\Delta V_{\rm th}[\phi(M_S)]
\ee
which, in our case, reduces to
\be
\label{masthreshold}
\Delta_{\rm th}m_H^2=
\frac{3}{4\pi^2}\frac{m_t^4(M_S)}{v^2(m_t)}
\frac{X_t^2}{M_S^2}\left(2-\frac{1}{6}\frac{X_t^2}{M_S^2}\right)
\ee
We have compared our analytical approximation~\cite{CEQW}
with the numerical NTLL result~\cite{CEQR} and found a difference
$\simlt 2$ GeV for all values of supersymmetric parameters.

The case $m_A<M_S$ is a bit more complicated since the effective theory
below the supersymmetric scale $M_S$ is the 2HDM. However since radiative
corrections in the 2HDM are equally dominated by the top-quark, we can
compute analytical expressions based upon the LL approximation
at the scale $Q^2\sim m_t^2$. This has been done in Ref.~\cite{CEQW}
where LL are resummed to two-loop. Our approximation differs from
the LL all-loop numerical resummation by $\simlt 3$ GeV, which we
consider the uncertainty inherent in the theoretical calculation,
provided the mixing is moderate and, in particular, bounded by the
condition,
\be
\label{condicion}
\left|\frac{m^2_{\;\widetilde{t}_1}-m^2_{\;\widetilde{t}_2}}
{m^2_{\;\widetilde{t}_1}+m^2_{\;\widetilde{t}_2}}\right|\simlt 0.5
\ee
where $\widetilde{t}_{1,2}$ are the two stop mass eigenstates.
In Fig.~5 the Higgs mass spectrum is plotted versus $m_A$.

\section{Threshold effects}

There are two possible caveats in the approximation we have just
presented: {\bf i)} Our expansion parameter $\log(M_S^2/m_t^2)$
does not behave properly in the supersymmetric limit $M_S\rightarrow 0$,
where we should recover the tree-level result. {\bf ii)} We have expanded
the threshold function $\Delta V_{\rm th}[\phi(M_S)]$ to order $X_t^4$.
In fact keeping the whole threshold function $\Delta V_{\rm th}[\phi(M_S)]$
we would be able to go to larger values of $X_t$ and to evaluate the
accuracy of the approximation (\ref{threshold}) and (\ref{masthreshold}).
Only then we will be able to
check the reliability of the maximum value of the
lightest Higgs mass (which corresponds to the maximal mixing) as provided
in the previous sections.
 %%%%%%%%%%%%%%%%%%%%%%%%%%%%%%%%%%%%%%%%%%%%%%%%%%%
\begin{figure}[htb]
%\psdraft
\centerline{
\psfig{figure=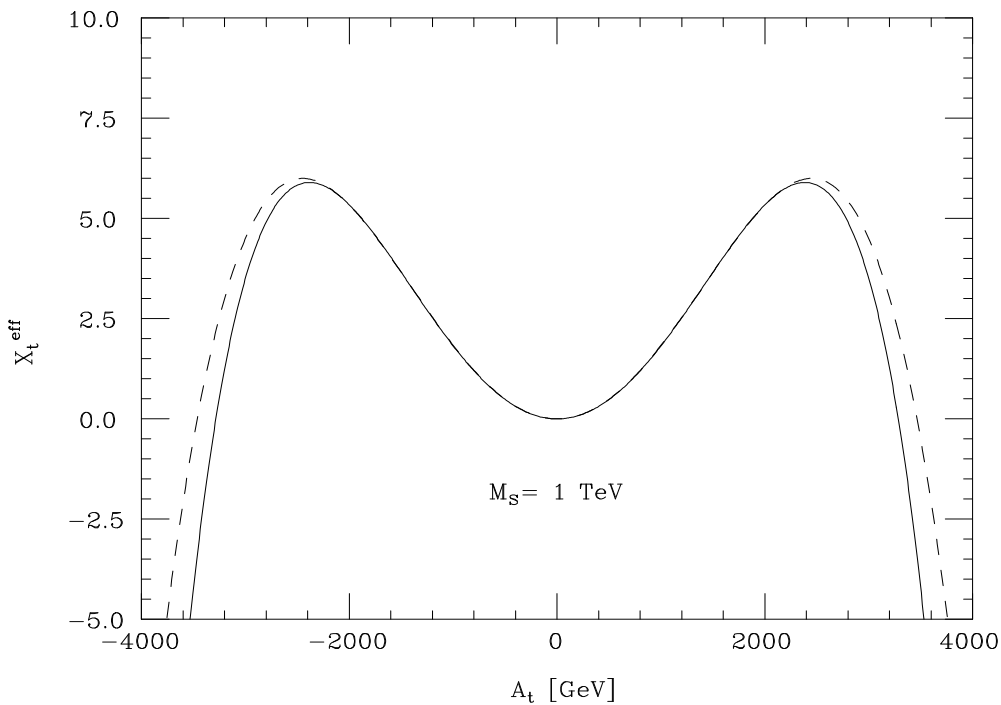,width=17cm,height=12cm,angle=90}}
\caption[0]{Plot of the exact (solid line) and approximated
(dashed line) effective mixing $X_t^{\rm eff}$ as a function of $A_t$,
for $M_S=1$ TeV and $\mu=0$.}
\end{figure}
%
%%%%%%%%%%%%%%%%%%%%%%%%%%%%%%%%%%%%%%%%%%%%%%%%%%%
%
This procedure has been properly followed in Refs.~\cite{CEQW} and
\cite{CQW}, where the most general case $m_Q\neq m_U\neq m_D$ has been
considered. We have proved that keeping the exact threshold function
$\Delta V_{\rm th}[\phi(M_S)]$, and properly running its value from the
high scale to $m_t$ with the corresponding anomalous dimensions as in
(\ref{Deltathm}), produces two effects: {\bf i)} It makes a resummation
from $M_S^2$ to $M_S^2+m_t^2$ and generates as (physical) expansion
parameter $\log[(M_S^2+m_t^2)/m_t^2]$. {\bf ii)} It generates a whole
threshold function $X_t^{\rm eff}$ such that (\ref{masthreshold})
becomes
\be
\label{masthreshold2}
\Delta_{\rm th}m_H^2=
\frac{3}{4\pi^2}\frac{m_t^4[M_S^2+m_t^2]}{v^2(m_t)}
X_t^{\rm eff}
\ee
and
\be
\label{desarrollo}
X_t^{\rm eff}=\frac{X_t^2}{M_S^2+m_t^2}
\left(2-\frac{1}{6}\frac{X_t^2}{M_S^2+m_t^2}\right)+\cdots
\ee
In fact we have plotted $X_t^{\rm eff}$ as a function of $A_t$ (solid line)
and compared with the approximation where we keep only terms up to
$X_t^4$ (dashed line), as we did in the previous sections.
The result shows that the
maximum of both curves are very close to each other, what justifies
the reliability of previous upper bounds on the lightest Higgs mass
as, e.g., in Fig.~3.

\section{Conclusions}

We have seen that effective potential methods, when decoupling is
properly accounted for, are useful and powerful techniques to analyze
the MSSM Higgs sector and provide easy-to-use analytical approximations
(with radiative corrections RG resummed)
to the Higgs mass spectrum and couplings.
In particular, an appropriate treatment of stop and sbottom decoupling
allows to consider large mixing parameters and put reliable upper
bounds on the lightest Higgs boson mass in the MSSM.

\end{document}